\documentclass[prd,twocolumn,showpacs,showkeys,footinbib]{revtex4}
\usepackage{amsmath}
\usepackage{graphics}

\renewcommand\vec[1]{\mathbf{#1}}
\newcommand\he[1]{#1^{\dagger}}
\DeclareMathOperator{\re}{Re}
\DeclareMathOperator{\im}{Im}
\DeclareMathOperator{\tr}{tr}

\begin{document}

\title{Goldstone boson counting in linear sigma models with chemical potential}
\author{Tom\'a\v s Brauner}
\email{brauner@ujf.cas.cz}
\affiliation{Department of Theoretical Physics, Nuclear Physics Institute, 25068 \v Re\v z, Czech Republic}

\begin{abstract}
We analyze the effects of finite chemical potential on spontaneous breaking of
internal symmetries within the class of relativistic field theories described
by the linear sigma model. Special attention is paid to the emergence of
``abnormal'' Goldstone bosons with quadratic dispersion relation. We show that
their presence is tightly connected to nonzero density of the Noether charges,
and formulate a general counting rule. The general results are demonstrated on
an $\mathrm{SU(3)}\times\mathrm{U(1)}$ invariant model with an
$\mathrm{SU(3)}$-sextet scalar field, which describes one of the
color-superconducting phases of QCD.
\end{abstract}

\maketitle

\section{Introduction}
Spontaneous symmetry breaking plays an important role in many areas of physics
and encounters a host of fascinating phenomena. The most distinguishing feature
of spontaneous symmetry breaking is the presence of soft modes, long-wavelength
fluctuations of the order parameter(s), guaranteed by the Goldstone theorem
\cite{Goldstone:1961eq,Goldstone:1962es}.

For low-energy properties of the spontaneously broken symmetry it is important
to know the number of the Goldstone bosons (GBs). While for spontaneously
broken internal symmetry (space-time symmetries will not be the subject of this
paper, see e.g. Ref. \cite{Low:2001bw}) in a Lorentz-invariant field theory it
is always equal to the number of broken symmetry generators, the original
Goldstone theorem predicts the existence of \emph{at least one} GB. Indeed,
there are several examples in nonrelativistic physics where the number of GBs
is smaller than one would naively expect. The most profound one is perhaps the
ferromagnet where the rotational $\mathrm{SO(3)}$ symmetry is spontaneously
broken down to $\mathrm{SO(2)}$, but only one GB (the magnon) exists.

The issue of GB counting in nonrelativistic field theories was enlightened by
Nielsen and Chadha \cite{Nielsen:1976hm}. They showed that the defect in the
number of GBs is related to the low-momentum behavior of their dispersion
relations. GBs with energy proportional to an odd power of momentum are
classified as type-I, and those with energy proportional to an even power of
momentum as type-II. The improved counting rule then states that \emph{the
number of GBs of type I plus twice the number of GBs of type II is greater or
equal to the number of broken generators}.

It should be noted that the form of the dispersion law of the lightest degrees
of freedom has important phenomenological consequences, e.g. for the
low-temperature thermodynamics of the system. For instance, the heat capacity
of a gas of bosons with $E\propto|\vec p|$ falls down as $T^3$ for $T\to0$,
while for bosons with $E\propto\vec p^2$ it is only $T^{3/2}$. If no
massless particles are present, the heat capacity is suppressed by factor
$e^{-m/kT}$, where $m$ is the mass of the lightest particle.

The interest in the problem of GB counting has been revived recently, mainly
thanks to the progress in understanding the phase diagram of quantum
chromodynamics. At finite density Lorentz invariance is explicitly broken and
GBs with nonlinear (as a matter of fact, generally quadratic) dispersion
relations may appear even in a relativistic field theory as a medium effect
\cite{Schafer:2001bq,Miransky:2001tw}. Their presence turns out to be connected
to the fact that some of the broken Noether charges develop nonzero density in
the ground state, as has been observed in various color-superconducting phases
of QCD \cite{Buballa:2002wy,Blaschke:2004cs} or in a neutron ferromagnet
\cite{Beraudo:2004zr}.

Schafer et al. \cite{Schafer:2001bq} have proved the following theorem: if the
commutators of all pairs of broken generators have zero ground-state
expectation value, then the number of GBs is equal to the number of broken
generators. It is therefore clear that the nonzero charge density itself is
not sufficient for a quadratic GB to appear. Indeed, the baryon number density
does not make any harm to the usual linear GBs in the color superconductors.
The corresponding generator must rather be a part of a non-Abelian symmetry
group. Our main goal is to show that the opposite to the theorem of Schafer et
al. generally holds: \emph{nonzero density of a commutator of two broken generators
implies one GB with quadratic dispersion law}.

The paper is organized as follows. The following section is devoted to preparatory
considerations: we explain how the quadratic GB is manifested in the Goldstone
commutator and sketch its realization in the linear sigma model. In the next part,
an example with an $\mathrm{SU(3)}$-sextet condensation is investigated in
detail. The general analysis is performed in the last section.

\section{Preliminary considerations}
In this section we shall investigate how the quadratic GBs come about, first at
the rather general level of the Goldstone commutator and later more explicitly
within the linear sigma model.

\subsection{Goldstone commutator}
Let us briefly recall the proof of the Goldstone theorem. Following Ref.
\cite{Nielsen:1976hm}, we assume there is a local (possibly composite) field
$\Phi(x)$ and a broken Noether charge $Q$ such that
$\langle0|[\Phi(x),Q]|0\rangle\neq0$. Inserting the complete set of
intermediate states into the commutator, one arrives at the representation
\begin{multline}
\langle0|[\Phi(x),Q]|0\rangle=\sum_{n=1}^l\left[e^{-iE_{\vec
k}t}\langle0|\Phi(0)|n_{\vec k}\rangle\langle n_{\vec
k}|j^0(0)|0\rangle\right.\\
\left.-e^{iE_{-\vec k}t}\langle0|j^0(0)|n_{-\vec k}\rangle\langle
n_{-\vec k}|\Phi(0)|0\rangle\right]\quad\text{at $\vec k=0$},
\label{eq:Goldstone_commutator}
\end{multline}
where the index $n$ counts the GBs.

Now assume that we deal with a non-Abelian symmetry group and some of its
charges have nonzero density in the ground state. Take as the GB field
$\Phi(x)$ the zero component of the Noether current itself, so that
$\langle0|[j^0_a(x),Q_b]|0\rangle=if_{abc}\langle0|j^0_c(x)|0\rangle$, where
$f_{abc}$ is the set of structure constants of the symmetry group. Should this
be nonzero, we infer from Eq. \eqref{eq:Goldstone_commutator} that both
$\langle0|j^0_a(0)|n\rangle$ and $\langle n|j^0_b(0)|0\rangle$ must be nonzero
for some Goldstone mode $n$.

The point of the above heuristic argument is that while in Lorentz-invariant
theories there is a one-to-one correspondence between the GBs and the broken
currents, here a single GB couples to two Noether currents. This explains (not
proves, of course) at a very elementary level how the GB counting rule is to be
modified in the presence of nonzero charge density.

One should perhaps note that the Nielsen--Chadha counting rule is formulated in
terms of the GB dispersion relations rather then charge densities. The
connection between these two was clarified by Leutwyler
\cite{Leutwyler:1994gf}, who showed by the analysis of the Ward identities for
the broken symmetry, that nonzero density of a non-Abelian charge induces a
term in the low-energy effective Lagrangian with a single time derivative. The
leading order effective Lagrangian is thus of the Schr\"odinger type and the
energy of the GB is proportional to momentum squared.

\subsection{Goldstone bosons within the linear sigma model}
In order to elaborate more on the properties of the GBs, we restrict ourselves
from now on to the framework of the linear sigma model, that is a general
scalar field theory with quartic self-interaction.

To see how the Goldstone commutator emerges in this language, recall the
$\mathrm{SU(2)}\times\mathrm{U(1)}$ invariant model
of Schafer et al. \cite{Schafer:2001bq} and Miransky and Shovkovy
\cite{Miransky:2001tw}. The Lagrangian for the complex doublet field $\phi$ of mass $M$
in Minkowski space reads
\begin{equation*}
\mathcal L=D_{\mu}\he\phi D^{\mu}\phi-M^2\he\phi\phi-\lambda(\he\phi\phi)^2.
\end{equation*}
Finite density of the statistical system is represented by the chemical
potential $\mu$, which enters the Lagrangian in terms of the covariant
derivative \cite{Kapusta:1981aa},
$D_{\mu}\phi=(\partial_{\mu}-i\delta_{0\mu}\mu)\phi$. Upon expanding the
covariant derivatives, the Lagrangian becomes
\begin{equation}
\mathcal
L=\partial_{\mu}\he\phi\partial^{\mu}\phi-2\mu\im\he\phi\partial_0\phi+
(\mu^2-M^2)\he\phi\phi-\lambda(\he\phi\phi)^2. \label{eq:Schafer_Lagrangian}
\end{equation}

For $\mu>M$ the static potential develops a nontrivial minimum and
the scalar field condenses. To find the spectrum of excitations at tree level
we reparameterize it as
\begin{equation*}
\phi=\frac1{\sqrt2}e^{i\pi_k\tau_k/v}\left(
\begin{array}{c}
0 \\ v+\varphi
\end{array}\right),\quad
v^2=\frac{\mu^2-M^2}{\lambda},
\end{equation*}
and look at the bilinear part of the Lagrangian. The crucial contribution comes
from the term in Eq. \eqref{eq:Schafer_Lagrangian} with one time derivative.
Upon expanding the exponentials it yields among others the expression
\begin{equation*}
-\frac12\mu\im\left(
\begin{array}{cc}
0 & 1
\end{array}\right)
[\pi_k\tau_k,\partial_0\pi_l\tau_l]\left(
\begin{array}{c}
0 \\ 1
\end{array}\right)=\mu(\pi_1\partial_0\pi_2-\pi_2\partial_0\pi_1).
\end{equation*}

As will be made clear in the next subsection, it is this term that is responsible
for the quadratic dispersion relation of one of the GBs. Its origin from the
nonzero density of a commutator of two generators is now made obvious. This is
the main idea to be remembered. The necessary technical details will come in
the next two sections.

\subsection{Bilinear Lagrangians and dispersion laws}
Bilinear Lagrangians with single-time-derivative terms will frequently occur
throughout the whole text. It is therefore worthwhile to fix once for all the
corresponding excitation spectrum.

The bilinear Lagrangians we will encounter will have the generic form
\begin{equation}
\mathcal
L_{\text{bilin}}=\frac12(\partial_{\mu}\pi)^2+\frac12(\partial_{\mu}H)^2-\frac12f^2(\mu)H^2-
g(\mu)H\partial_0\pi. \label{eq:bilinear_Lagrangian}
\end{equation}
The notation suggests that $H$ is a massive (Higgs) mode whose mass function
$f^2(\mu)$ depends on the chemical potential, while $\pi$ is the Goldstone
mode. The excitation spectrum is found from the poles of the two-point Green
functions or, equivalently, by solving the condition
\begin{equation*}
\det\left(
\begin{array}{cc}
E^2-\vec p^2 & +iEg(\mu)\\
-iEg(\mu) & E^2-\vec p^2-f^2(\mu)
\end{array}\right)=0.
\end{equation*}
It turns out there is one massive mode, with dispersion relation
\begin{equation}
E^2=f^2(\mu)+g^2(\mu)+\mathcal O(\vec p^2),
\label{eq:disp_rel_massive}
\end{equation}
and one massless mode, with dispersion relation
\begin{equation}
E^2=\frac{f^2(\mu)}{f^2(\mu)+g^2(\mu)}\vec
p^2+\frac{g^4(\mu)}{\left[f^2(\mu)+g^2(\mu)\right]^3}\vec p^4+\mathcal O(\vec p^6).
\label{eq:disp_rel_massless}
\end{equation}

Now if $f^2(\mu)>0$, the Lagrangian \eqref{eq:bilinear_Lagrangian} indeed
describes a massive particle and a GB, whose energy is linear in momentum in
the long-wavelength limit. On the other hand, when $f^2(\mu)=0$, that is when
both $\pi$ and $H$ would correspond to linear GBs in the absence of the
chemical potential, the dispersion relation of the gapless mode reduces to
$E=\vec p^2/|g(\mu)|$. This is the sought quadratic Goldstone.

In conclusion, the term with a single time derivative in general mixes the
original fields in the Lagrangian. Mixing of a massive mode with a massless one
yields one massive particle and one linear GB, mixing of two massless modes
results in a massive particle and a quadratic GB \footnote{There is an exception to the
first case: right at the phase transition point, $f^2(\mu)=0$ even for the Higgs mode,
the phase velocity of the linear GB goes to zero, and its dispersion relation
becomes quadratic \cite{Schafer:2001bq}.}.

\section{Linear sigma model for $\mathrm{SU(3)}$-sextet condensation}
As a nontrivial demonstration of the general idea proposed in the previous
section, we shall now analyze in detail a particular model of spontaneous
symmetry breaking. Consider a scalar field $\Phi$ that transforms as a
symmetric rank-two tensor under the group $\mathrm{SU(3)}$, $\Phi\to U\Phi
U^T$. Such a field describes a one-flavor diquark condensate in one of the
superconducting phases of QCD \cite{Brauner:2003pj}.

In addition to the $\mathrm{SU(3)}$ group, $\Phi$ is subject to $\mathrm{U(1)}$
transformations corresponding to quark number,
$\Phi\to e^{i\theta}\Phi e^{i\theta}=e^{2i\theta}\Phi$. The most
general $\mathrm{SU(3)}\times\mathrm{U(1)}$ invariant Lagrangian has the form
\begin{equation}
\mathcal L=\tr(D_{\mu}\he\Phi
D^{\mu}\Phi)-M^2\tr\he\Phi\Phi-a\tr(\he\Phi\Phi)^2-b(\tr\he\Phi\Phi)^2.
\label{eq:sextet_Lagrangian}
\end{equation}
The quark-number $\mathrm{U(1)}$ has been assigned chemical potential $\mu$ so
that $D_0\Phi=(\partial_0-2i\mu)\Phi$. The parameters $a,b$ are constrained by
the requirement of boundedness of the static potential \cite{Brauner:2003pj}.
It is necessary that either both are non-negative (and at least one of them
nonzero), or $a<0$ and $b>|a|$, or $b<0$ and $a>3|b|$.

\subsection{Minimum of the static potential}
We start our analysis with a careful inspection of the static potential,
\begin{equation}
V(\Phi)=-(4\mu^2-M^2)\tr\he\Phi\Phi+a\tr(\he\Phi\Phi)^2+b(\tr\he\Phi\Phi)^2.
\label{eq:sextet_potential}
\end{equation}
A potential of the same type has been analyzed by Iida and Baym
\cite{Iida:2000ha}. In their case, however, the global symmetry was different,
and we therefore provide full details.

When $4\mu^2-M^2>0$, the stationary point $\Phi=0$ becomes unstable and a new,
nontrivial minimum appears \footnote{It should be stressed that our analysis applies
to both signs of $M^2$. For $M^2>0$ the model describes relativistic Bose--Einstein
condensation, which occurs at $2\mu>M$. For $M^2<0$ the Lagrangian \eqref{eq:sextet_Lagrangian}
represents simply a spontaneously broken symmetry at finite density.}. The stationary-point
condition reads
\begin{equation}
\Phi\left(-4\mu^2+M^2+2a\he\Phi\Phi+2b\tr\he\Phi\Phi\right)=0.
\label{eq:stat_point_cond}
\end{equation}

Before going into detailed solution of this equation we note that by
multiplying Eq. \eqref{eq:stat_point_cond} from left by $\he\Phi$ and taking
the trace, the stationary-point value of the potential
\eqref{eq:sextet_potential} is found to be
\begin{equation*}
V_{\text{stat}}=-\frac12(4\mu^2-M^2)\tr\he\Phi\Phi.
\end{equation*}
Any nontrivial stationary point of the potential is thus energetically more
favorable than the perturbative vacuum $\Phi=0$. We are, however, obliged to
find a stable ground state, that is the absolute minimum of the potential.

We now make use of the fact that the field $\Phi$ can always
be brought by a suitable $\mathrm{SU(3)}\times\mathrm{U(1)}$ transformation to
the standard form, which is a real diagonal matrix with non-negative entries \cite{Schur:1945ab}.
Eq. \eqref{eq:stat_point_cond} then splits into three conditions and it is easy
to see that all nonzero diagonal elements acquire the same value, denoted here
by $\Delta$.

Let there be $n$ of them, $n=1,2,3$. Eq. \eqref{eq:stat_point_cond} implies
\begin{equation*}
\Delta^2=\frac12\frac{4\mu^2-M^2}{a+bn},\quad
V_{\text{stat}}=-\frac14\frac{(4\mu^2-M^2)^2}{b+\frac an}.
\end{equation*}
To find the absolute minimum of the potential, it remains to minimize this
expression with respect to $n$.

For $a>0$ the minimum occurs at $n=3$, and $\Phi$ is proportional to the unit
matrix, $\Phi=\Delta\openone$, where
\begin{equation*}
\Delta^2=\frac12\frac{4\mu^2-M^2}{a+3b}.
\end{equation*}
The $\mathrm{SU(3)}\times\mathrm{U(1)}$ symmetry
is broken down to $\mathrm{SO(3)}$.

For $a<0$ the potential is minimized by
$n=1$, that is $\Phi$ is diagonal with a single nonzero entry and is
conventionally chosen to be $\Phi=\mathrm{diag}(0,0,\Delta)$, where now
\begin{equation*}
\Delta^2=\frac12\frac{4\mu^2-M^2}{a+b}.
\end{equation*}
The unbroken subgroup is now $\mathrm{SU(2)}\times\mathrm{U(1)}$.

For $a=0$ the local minima corresponding to different $n$ are degenerate since
in that case, the Lagrangian \eqref{eq:sextet_Lagrangian} is invariant under an
enhanced $\mathrm{SU(6)}\times\mathrm{U(1)}$ symmetry, treating $\Phi$ as a
fundamental sextet. Nonzero ground-state expectation value of $\Phi$ breaks
this symmetry to $\mathrm{SU(5)}\times\mathrm{U(1)}$. As we shall see, such an
enhanced symmetry leads to an increased number of GBs with quadratic dispersion
relation \cite{Sannino:2001fd}.

\subsection{Noether currents and charge densities}
Having found the vacuum configuration of the scalar field, we are ready to
reparameterize it and find the excitation spectrum from the bilinear part
of the Lagrangian. Before doing that, we evaluate the ground-state densities of
the Noether charges in order to make \emph{a priori} predictions about the nature of
the GBs.

The infinitesimal $\mathrm{SU(3)}\times\mathrm{U(1)}$ transformation of $\Phi$
has the generic form $\delta\Phi=i\theta_k(\lambda_k\Phi+\Phi \lambda_k^T)$,
where the $\lambda_k$ stands for the Gell-Mann matrices ($k=1,\dotsc,8$) and
the unit matrix ($k=0$), respectively. The corresponding Noether currents are
\begin{equation*}
j^{\mu}_k=-i\tr\left[D^{\mu}\he\Phi\left(\lambda_k\Phi+\Phi
\lambda_k^T\right)-\mathrm{h.c.}\right].
\end{equation*}
Taking a generic static field configuration to be
$\Phi=\mathrm{diag}(\Delta_1,\Delta_2,\Delta_3)$ results in the charge
densities
\begin{equation*}
\begin{split}
j^0_0&=8\mu(\Delta_1^2+\Delta_2^2+\Delta_3^2),\\
j^0_3&=8\mu(\Delta_1^2-\Delta_2^2),\\
j^0_8&=\frac8{\sqrt3}\mu(\Delta_1^2+\Delta_2^2-2\Delta_3^2).
\end{split}
\end{equation*}

In the $\mathrm{SO(3)}$ symmetric phase ($a>0$), all generators but the
$\mathrm{U(1)}$ quark number have zero density. As this is an Abelian
generator, we expect six linear GBs corresponding to the six broken generators
$\openone,\lambda_1,\lambda_3,\lambda_4,\lambda_6,\lambda_8$. In the $a<0$
case, the densities of $\lambda_0$ and $\lambda_8$ are nonzero. This means that
the commutators $[\lambda_4,\lambda_5]$ and $[\lambda_6,\lambda_7]$ have
nonzero ground-state density. With regard to the general discussion above, we
thus expect two quadratic GBs corresponding to pairs $(\lambda_4,\lambda_5)$
and $(\lambda_6,\lambda_7)$, and one linear GB of the generator $\lambda_8$.

\subsection{The $a>0$ case}
We shall now proceed to the calculation of the mass spectrum of the $a>0$
phase. We could do well with just shifting $\Phi$ by its vacuum expectation
value $\Delta\openone$, but this would complicate the identification of the
massless modes. It is more convenient, and physical, to find such a
parameterization that the GBs disappear from the static potential.

To that end, recall that the field $\Phi(x)$ (now coordinate-dependent) can be
brought to the diagonal form by a suitable $\mathrm{SU(3)}\times\mathrm{U(1)}$
transformation. In other words, it may be written as
\begin{equation*}
\Phi(x)=e^{2i\theta(x)}U(x)D(x)U^T(x),
\end{equation*}
where $U(x)\in\mathrm{SU(3)}$ and $D(x)$ is real, diagonal and non-negative.
Now the unitary matrix $U$ can be (at least in the vicinity of unity) expressed
as a product $U=VO$, $O\in\mathrm{SO(3)}$ being an element of the unbroken
subgroup and $V$ being built from the broken generators,
$V=e^{i\pi_k\lambda_k}$, $k=1,3,4,6,8$. A simple observation that
$O(x)D(x)O^T(x)$ is the general parameterization of a real symmetric matrix
leads to the final prescription,
\begin{equation*}
\Phi(x)=e^{2i\theta(x)}V(x)\left[\Delta\openone+\varphi(x)\right]V^T(x).
\end{equation*}
The real symmetric matrix $\varphi$ contains six massive modes, while $V$
contains five GBs. With $\theta$ this is altogether twelve degrees of freedom,
as it should for $\Phi$ is a complex symmetric $3\times3$ matrix.

It is now straightforward, though somewhat tedious, to plug this
parameterization into the Lagrangian \eqref{eq:sextet_Lagrangian} and expand to
the second order in the fields. Omitting details of the calculations, we just
report on the results.

The full static potential (up to a constant term -- the vacuum energy density)
becomes
\begin{multline*}
V(\Phi)=4\Delta^2\left[a\tr\varphi^2+b(\tr\varphi)^2\right]\\
+4\Delta\left(a\tr\varphi^3+b\tr\varphi\tr\varphi^2\right)+a\tr\varphi^4+b(\tr\varphi^2)^2.
\end{multline*}
The bilinear Lagrangian turns out to be (we use the notation $V=e^{i\Pi}$)
\begin{multline*}
\mathcal
L_{\text{bilin}}=12\Delta^2(\partial_{\mu}\theta)^2+4\Delta^2\tr(\partial_{\mu}\Pi)^2
+\tr(\partial_{\mu}\varphi)^2\\
-4\Delta^2\left[a\tr\varphi^2+b(\tr\varphi)^2\right]
-16\mu\Delta\left[\partial_0\theta\tr\varphi+\tr(\varphi\partial_0\Pi)\right].
\end{multline*}
The kinetic terms are brought to the canonical form by a simple rescaling of the
fields, upon which the spectrum is readily determined from Eqs.
\eqref{eq:disp_rel_massive} and \eqref{eq:disp_rel_massless}.

The excitations fall into irreducible multiplets of the unbroken
$\mathrm{SO(3)}$ group. There are two singlets, stemming from the mixing of
$\theta$ and $\tr\varphi$,
\begin{align*}
\text{massive mode} && E^2&=24\mu^2-2M^2+\mathcal O(\vec p^2),\\
\text{linear GB} && E^2&=\frac{4\mu^2-M^2}{12\mu^2-M^2}\vec
p^2+\mathcal O(\vec p^4),
\end{align*}
and two $5$-plets, the mixtures of ($\pi_1,\pi_3,\pi_4,\pi_6,\pi_8$) and the
traceless part of $\varphi$,
\begin{align*}
\text{massive modes} &&
E^2&=\frac{(24\mu^2-2M^2)a+48\mu^2b}{a+3b}+\mathcal O(\vec p^2),\\
\text{linear GBs} &&
E^2&=\frac{(4\mu^2-M^2)a}{(12\mu^2-M^2)a+24\mu^2b}\vec p^2+\mathcal O(\vec
p^4).
\end{align*}
It is easily seen from these formulas that the masses of the massive singlet and
the massive $5$-plet are connected by
\begin{equation*}
m_{\mathbf 1}^2=m_{\mathbf 5}^2+(4\mu^2-M^2)\frac{6b}{a+3b}=m_{\mathbf
5}^2+12\Delta^2b.
\end{equation*}
The singlet is heavier than the $5$-plet for $b>0$ and vice versa.

\begin{figure}[t]
\begin{center}
\framebox{\scalebox{0.85}{\include{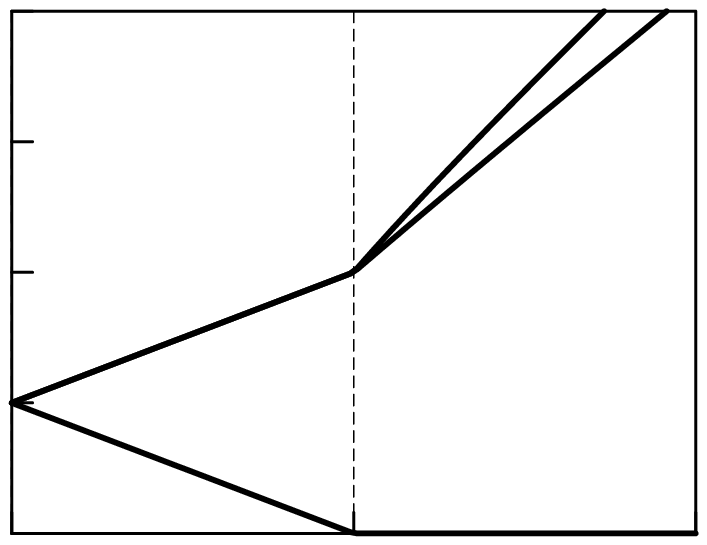}}}
\caption{Mass spectrum as a function of the chemical potential for $a>0$.
The boldface-typed numbers denote
the degeneracies of the excitation branches. To obtain numerical results,
particular values $a=b=1$ were chosen.} \label{fig:spectrum1}
\end{center}
\end{figure}

The excitation spectrum is plotted in Fig. \ref{fig:spectrum1} for the case
$M^2>0$. Below the phase transition to the Bose--Einstein-condensed phase, the
medium-modified dispersion relations are simply $E=\sqrt{\vec p^2+M^2}\pm2\mu$.
Right at the transition point, there are six modes with mass $2M$ and six
massless ones with dispersion $E=\vec p^2/4\mu$. As the phase transition is
second order, the dispersion relations of all excitation branches must be
continuous functions of $\mu$, that is \emph{all GBs become quadratic at the
transition point}. This is also easily checked on the broken-symmetry side of
the transition. As $2\mu\to M+$, the phase velocities of the linear GBs tend to
zero, and their dispersions become quadratic.

Note that also for $a=0$ the dispersion relation of the GB $5$-plet becomes
quadratic, $E=\vec p^2/4\mu$. This is in accord with the enhanced
$\mathrm{SU(6)\times\mathrm{U(1)}}$ symmetry of the Lagrangian. There are
altogether eleven broken generators of the coset $\mathrm{SU(6)/SU(5)}$, one
linear GB and five quadratic ones [forming now the $5$-plet of the unbroken
$\mathrm{SU(5)}$], and the Nielsen--Chadha counting rule is thus satisfied.

\subsection{The $a<0$ case}
We use the same method for parameterization of $\Phi$ as in the previous case.
This time we write $\Phi(x)=U(x)D(x)U^T(x)$, where
$U(x)\in\mathrm{SU(3)}\times\mathrm{U(1)}$. Next perform the decomposition
$U=e^{i\Pi}U'$, where $\Pi=\pi_k\lambda_k$, $k=4,5,6,7,8$, and $U'$
belongs to the unbroken subgroup $\mathrm{SU(2)}\times\mathrm{U(1)}$.
Since $U'(x)D(x)U'^T(x)$ is block-diagonal with a complex symmetric $2\times2$
matrix in the upper-left corner, we arrive at the parameterization
\begin{widetext}
\begin{equation*}
\Phi(x)=e^{i\Pi(x)}\left[\mathrm{diag}(0,0,\Delta)+\Sigma(x)\right]
e^{i\Pi^T(x)},\quad
\Sigma(x)=\left(
\begin{array}{cc|c}
\sigma(x) & & \\
 & & \\
\hline
\phantom{H(x)} & \phantom{H(x)} & H(x)
\end{array}\right).
\end{equation*}
Here $H$ is a real field and $\sigma$ is a complex symmetric $2\times2$ matrix.
These two embody the massive modes that survive in the static potential,
\begin{equation*}
V(\Phi)=4\Delta^2(a+b)H^2-2\Delta^2a\tr\he\sigma\sigma
+4\Delta(a+b)H^3+(a+b)H^4+4\Delta bH\tr\he\sigma\sigma
+2bH^2\tr\he\sigma\sigma+a\tr(\he\sigma\sigma)^2+
b(\tr\he\sigma\sigma)^2.
\end{equation*}
The bilinear part of the Lagrangian reads
\begin{multline*}
\mathcal
L_{\text{bilin}}=\tr(\partial_{\mu}\he\sigma\partial^{\mu}\sigma)+(\partial_{\mu}H)^2+
2\Delta^2(\partial_{\mu}\Pi\partial^{\mu}\Pi)_{33}+2\Delta^2(\partial_{\mu}\Pi_{33})^2-
4\Delta^2(a+b)H^2+2\Delta^2a\tr\he\sigma\sigma\\
-16\mu\Delta H\partial_0\Pi_{33}-
4\mu\Delta^2\im[\Pi,\partial_0\Pi]_{33}-4\mu\im\tr\he\sigma\partial_0\sigma.
\end{multline*}
\end{widetext}

The excitations are again organized in multiplets of the unbroken
$\mathrm{SU(2)}\times\mathrm{U(1)}$. $H$ and $\pi_8$ mix to form two singlets,
\begin{align*}
\text{massive mode} && E^2&=24\mu^2-2M^2+\mathcal O(\vec p^2),\\
\text{linear GB} && E^2&=\frac{4\mu^2-M^2}{12\mu^2-M^2}\vec
p^2+\mathcal O(\vec p^4),
\end{align*}
and the pairs $(\pi_4,\pi_5)$ and $(\pi_6,\pi_7)$ give rise to a doublet of
massive modes and a doublet of massless ones,
\begin{align*}
\text{massive modes} &&
E^2&=16\mu^2+\mathcal O(\vec p^2),\\
\text{quadratic GBs} && E^2&=\frac{\vec p^4}{16\mu^2}+\mathcal O(\vec p^6).
\end{align*}

The matrix $\sigma$ represents two triplets of massive particles. The part of
the bilinear Lagrangian containing $\sigma$ may be rewritten as
\begin{equation*}
\mathcal L_{\sigma}=\tr(D_{\mu}\he\sigma
D^{\mu}\sigma)-(4\mu^2+2\Delta^2|a|)\tr\he\sigma\sigma,
\end{equation*}
which immediately implies the dispersion relations
\begin{equation*}
E=\sqrt{4\mu^2+2\Delta^2|a|}\pm2\mu+\mathcal O(\vec p^2).
\end{equation*}

\begin{figure}[t]
\begin{center}
\framebox{\scalebox{0.85}{\include{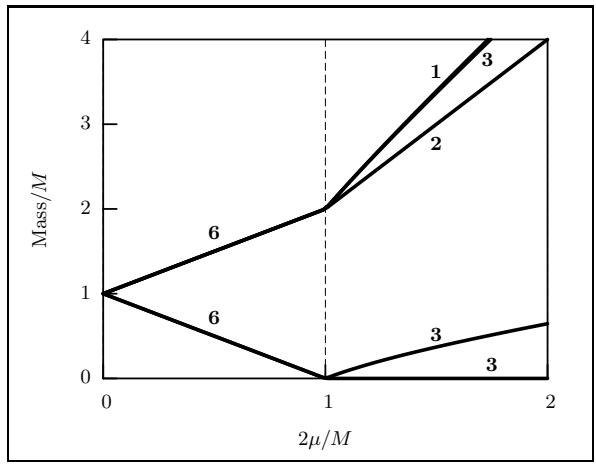}}}
\caption{Mass spectrum as a
function of the chemical potential for $a<0$. The singlet and triplet lines are
so close that they almost coincide, but they are not degenerate. The spectrum
is plotted for $a=-0.5$ and $b=1$.}
\label{fig:spectrum2}
\end{center}
\end{figure}

The mass spectrum is shown in Fig. \ref{fig:spectrum2}. The unbroken-phase part
of the spectrum is the same as in the $a>0$ case, since for $2\mu<M$ the
\emph{tree-level} masses of the particles do not depend at all on the quartic
potential, i.e. the parameters $a,b$. Also, the same remark about the
continuity of the dispersion relations across the phase transition applies.

Again, in the limit $a=0$, the lighter of the two triplets in $\sigma$ becomes
a triplet of quadratic GBs, and joins the other two quadratic GBs to form the
full $\mathrm{SU(5)}$ 5-plet.

\begin{figure}[t]
\begin{center}
\framebox{\scalebox{0.85}{\include{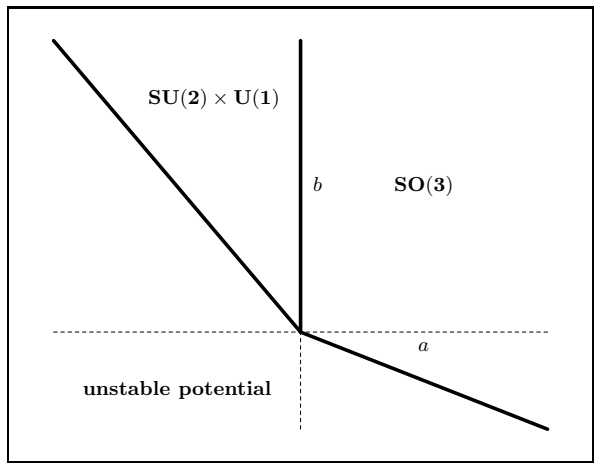}}}
\caption{Phase diagram of
the linear sigma model for $\mathrm{SU(3)}$-sextet condensation. The phases are
labeled by the symmetry of the ground state. The line of second order phase
transition at $a=0,b>0$ has $\mathrm{SU(5)} \times\mathrm{U(1)}$ symmetry.}
\label{fig:phase_diagram}
\end{center}
\end{figure}

To summarize our results, the theory described by the Lagrangian
\eqref{eq:sextet_Lagrangian} has two different ordered phases, both occurring
at $4\mu^2>M^2$, distinguished by the symmetry of the ground state. The
corresponding phase diagram in the $(a,b)$ plane is displayed in Fig.
\ref{fig:phase_diagram}.

\begin{figure}[t]
\begin{center}
\framebox{\scalebox{0.85}{\include{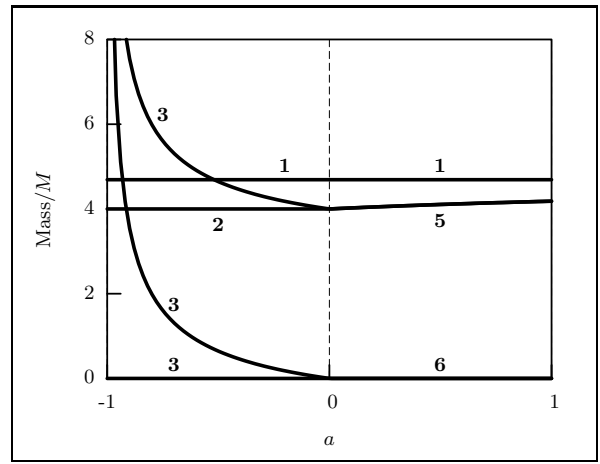}}}
\caption{Mass spectrum as a function of $a$. The graph is plotted for $\mu=M$
and $b=1$. The potential is unstable for $a<-1$. The singular behavior of the
masses of $\sigma$ is due to divergence of $\Delta$ towards the stability limit
of the potential.}
\label{fig:phase_transition}
\end{center}
\end{figure}

As the excitations above the ordered ground state are grouped into irreducible
multiplets of the unbroken symmetry, it is interesting to find out how the
structure of these multiplets changes across the phase transition from one
ordered phase to the other. In Fig. \ref{fig:phase_transition} we show the
dependence of the masses on the parameter $a$ at constant chemical potential.
The masses are continuous functions of $a$ as the transition is second order.

As a final remark we note that in the original application of Ref.
\cite{Brauner:2003pj}, the field $\Phi$ represented a diquark condensate and
the $\mathrm{SU(3)}$ was the color gauge group of QCD. One might wonder whether
the usual Higgs mechanism for gauge boson masses survives when there are fewer
GBs than the number of broken generators, because of the presence of quadratic
GBs. This question was answered affirmatively by Gusynin et al.
\cite{Gusynin:2003yu}, and there is therefore no need to worry about the fate
of gluons.

\section{General analysis}
In this section we shall collect experience gained by solving particular
examples and set out for a general analysis. We will find out, with some
effort, that the ideas sketched in Sec. II and demonstrated in Sec. III have a
straightforward generalization to a whole class of theories. It is understood,
however, that we shall all the time stay in the framework of the linear sigma
model, and at the tree level. The possibilities of further progress are
discussed in the conclusions.

\subsection{Chemical potential and global symmetry}
As the starting point we shall address the question what is the most general
symmetry of a theory with nonzero chemical potential.

Let the microscopic theory possess a global continuous symmetry with the
corresponding conserved Noether charges. The physical meaning of the chemical
potential $\mu$ is that we wish to fix the statistical average of a conserved
charge, say $Q$. This is technically achieved by introducing the grandcanonical
ensemble and replacing the microscopic Hamiltonian $H$ with $H-\mu Q$.

It is now clear that by adding the chemical potential, we break explicitly all
Noether charges that do not commute with $Q$. This is the technical realization
of the physically intuitive fact that we cannot keep simultaneously fixed the
values of two noncommuting operators (i.e. incompatible observables).

This simple observation implies that, as far as \emph{exact} symmetry is
concerned, chemical potential is always assigned to a generator that commutes
with all others, that is to a $\mathrm{U(1)}$ factor of the exact global
symmetry group.

Of course, when the symmetry of the microscopic theory is non-Abelian, then
adding of the chemical potential generally produces a number of approximately
conserved charges (at least for small $\mu$) that generate approximate
symmetries. These may also be spontaneously broken, resulting in the
corresponding set of pseudo-Goldstone bosons. Throughout this paper we are,
however, concerned only with true GBs, and therefore only the exact global
symmetry will be considered.

It is also interesting to find out how the Abelian nature of the charge
equipped with chemical potential is manifested in the Lagrangian formalism.
There, as already mentioned, chemical potential enters the Lagrangian in terms
of the covariant derivative of ``matter'' fields \cite{Kapusta:1981aa}.

The Lagrangian can be made formally \emph{gauge-invariant} by introducing an
external gauge field $A_{\mu}$. Provided the matter fields $\phi$ transform
under the symmetry group linearly as $\phi\to U\phi$, $A_{\mu}$ transforms as
usual as $A_{\mu}\to UA_{\mu}U^{-1}+iU\partial_{\mu}U^{-1}$. Now the exact
symmetry is such that the Lagrangian is invariant under the \emph{global}
transformation of the matter fields with $A_{\mu}$ \emph{fixed} at
$A_{\mu}=(\mu Q,0,0,0)$. This is possible only when $A_{\mu}=UA_{\mu} U^{-1}$.
We thus again arrive at the conclusion that the generator being assigned
chemical potential must commute with all others.

\subsection{Linear sigma model}
Now consider a general linear sigma model defined by the Lagrangian
\begin{equation}
\mathcal L=D_{\mu}\he\phi D^{\mu}\phi-V(\phi).
\label{eq:Lagrangian_sigma_model}
\end{equation}
Here $\phi$ denotes a set of complex \footnote{Real representations are not
interesting for us as they cannot give rise to nonzero charge density: $\he\phi
Q\phi=0$ simply because of the antisymmetry of the charge matrix $Q$.} scalar
fields that form a (possibly reducible) multiplet of the exact global symmetry
group $\mathrm{G}$, i.e. span the target space of a (possibly reducible)
representation of $\mathrm{G}$, say $\mathcal R$. $V(\phi)$ is the most general
$\mathrm{G}$-invariant static potential containing terms up to the fourth power
of $\phi$, and the covariant derivative is given by
$D_{\mu}\phi=(\partial_{\mu}-iA_{\mu})\phi$. $A_{\mu}$ is the constant external
field that incorporates chemical potential for one or more $\mathrm{U(1)}$
factors of $\mathrm{G}$, and is eventually set to
$A_{\mu}=(\sum_i\mu_iQ_{i\mathcal R} ,0,0,0)$, where the $Q_i$'s are the
$\mathrm{U(1)}$ generators, the subscript $\mathcal R$ denoting the image in
the representation $\mathcal R$.

Upon expanding the covariant derivatives Eq. \eqref{eq:Lagrangian_sigma_model}
takes the form
\begin{equation}
\mathcal L=\partial_{\mu}\he\phi\partial^{\mu}\phi-2\im\he\phi
A^{\mu}\partial_{\mu}\phi-V_{\text{eff}}(\phi),
\label{eq:Lagrangian_sigma_model_expanded}
\end{equation}
the effective $\mu$-dependent potential being
$V_{\text{eff}}(\phi)=V(\phi)-\he\phi A^{\mu}A_{\mu}\phi$.

Spontaneous symmetry breaking occurs when $V_{\text{eff}}(\phi)$ develops a
nontrivial minimum at some $\phi=\phi_0$. In order to elucidate the physical
content of such a theory, it is necessary to conveniently parameterize the
field $\phi$.

We stress the generality of the parameterization method suggested and applied
in Sec. III. One first writes $\phi(x)=U_{\mathcal R}(x)\phi_{\text{std}}(x)$,
where $\phi_{\text{std}}$ is a standard form to which the field $\phi$ can
always be brought by a suitable transformation $U\in\mathrm{G}$. Next
$U_{\mathcal R}$ is factorized as $U_{\mathcal R}=e^{i\Pi}U'_{\mathcal R}$,
where $\Pi$ is a linear combination of the broken generators (or more
precisely, their $\mathcal R$-images) and $U'$ belongs to the unbroken subgroup
$\mathrm{H}$. The final step is to identify $U'_{\mathcal
R}(x)\phi_{\text{std}}(x)$ with a certain representation of $\mathrm{H}$ and
parameterize it linearly as $\phi_0+H(x)$. $H(x)$ is going to be the multiplet
of massive (Higgs) fields. We therefore invoke the parameterization
\begin{equation}
\phi(x)=e^{i\Pi(x)}\left[\phi_0+H(x)\right].
\label{eq:parameterization_nonlinear}
\end{equation}

In order to specify the transformation properties of $H$, recall that the GBs
transform linearly in the adjoint representation of the unbroken subgroup
\cite{Coleman:1969sm}, i.e. $\Pi\to U'_{\mathcal R}\Pi U'^{-1}_{\mathcal R}$
for any $U'\in\mathrm H$. As a consequence, $H=e^{-i\Pi}\phi-\phi_0$ transforms
as $H\to U'_{\mathcal R}H$, since $\phi_0$ is an $\mathrm H$-singlet.

To summarize, $H$ transforms in the representation $\mathcal R$ truncated to
the subgroup $\mathrm H$, and the multiplets of the massive modes are therefore
found in the decomposition of $\mathcal R$ into irreducible representations of
$\mathrm H$.

For instance, in our case $a>0$ the symmetric rank-two tensor representation of
$\mathrm{SU(3)}$ splits under the $\mathrm{SO(3)}$ subgroup into a traceless
symmetric rank-two tensor and a singlet. On the other hand, in the $a<0$ case
it yields a symmetric rank-two tensor of $\mathrm{SU(2)}$ (the field $\sigma$)
plus a singlet.

As an aside let us remark that the physical spectrum of the theory of course
does not depend on the parameterization chosen for the field $\phi$. What if we
chose e.g. the linear parameterization mentioned (and abandoned) above in Sec
III C? Instead of Eq. \eqref{eq:parameterization_nonlinear}, we would then have
analogously
\begin{equation}
\phi(x)=\phi_0+H(x)+i\Pi(x)\phi_0. \label{eq:parameterization_linear}
\end{equation}

It is easy to see that the \emph{bilinear} terms in the Lagrangian with one or
two derivatives come out identical as for the parameterization
\eqref{eq:parameterization_nonlinear}. The reason is that the only difference
stemming from the nonlinear structure of $e^{i\Pi}$ could possibly come in the
form $\he\phi_0A^{\mu}\partial_{\mu}\Pi^2\phi_0$, but this is real and
therefore it drops out of the Lagrangian
\eqref{eq:Lagrangian_sigma_model_expanded}.

The only difficulty with the linear parameterization
\eqref{eq:parameterization_linear} is that the GBs do not disappear
automatically from the static potential. Instead, we have to use explicitly the
$\mathrm G$-invariance to show that $\Pi$ disappears from the \emph{bilinear}
(mass) part of the potential.

Upon the field redefinition as in Eq. \eqref{eq:parameterization_nonlinear},
the effective potential $V_{\text{eff}}$ becomes (up to a constant term)
\begin{equation*}
V_{\text{eff}}(\phi)=V(\phi_0+H)-(\he HA^{\mu}A_{\mu}H+2\re\he
HA^{\mu}A_{\mu}\phi_0).
\end{equation*}
As we are expanding the potential about its absolute minimum, the additional
term linear in $H$ is right enough to cancel a similar term coming from
$V(\phi_0+H)$. We are interested in the bilinear part of the potential,
$V_{\text{bilin}}(H)$, which determines the mass term for $H$.

Now we analyze the first two terms of the Lagrangian
\eqref{eq:Lagrangian_sigma_model_expanded}. The two-derivative term yields the
bilinear contribution
\begin{equation}
\partial_{\mu}\he
H\partial^{\mu}H+\he\phi_0\partial_{\mu}\Pi\partial^{\mu}\Pi\phi_0+2\im\he\phi_0\partial_{\mu}\Pi\partial^{\mu}H.
\label{eq:two_derivative}
\end{equation}
The first two terms in Eq. \eqref{eq:two_derivative} are the expected kinetic
terms for the Higgs and Goldstone fields, respectively. The GB term, however,
asks for a check that it is nondegenerate.

Let $\Pi(x)=\pi_k(x)T_k$, $T_k$ being the set of broken generators. The GB
kinetic term becomes
$\partial_{\mu}\pi_k\partial^{\mu}\pi_l\he\phi_0T_kT_l\phi_0=\frac12\partial_{\mu}\pi_k\partial^{\mu}
\pi_l\he\phi_0\{T_k,T_l\}\phi_0$. The matrix $\he\phi_0\{T_k,T_l\}\phi_0$ is
real and symmetric and may be chosen, by taking an appropriate basis of broken
generators, diagonal. It is obviously nondegenerate, as necessary in order to
have kinetic terms for all the GBs, since otherwise $\he\phi_0T_kT_k\phi_0=0$
for some $T_k$, implying that $T_k$ is in fact not broken.

The third term in Eq. \eqref{eq:two_derivative} eventually turns out to be
zero. Nevertheless, as other terms of a similar structure will be dealt with in
the following, we shall analyze it in detail. The crucial point is the way
various fields transform under the unbroken subgroup $\mathrm H$. Virtually all
information about the structure of the bilinear Lagrangian may be obtained by a
proper decomposition of the representation $\mathcal R$ into irreducible
representations of $\mathrm H$, and making repeated use of the Wigner--Eckart
theorem.

Now when $H$ and $\Pi$ belong to different representations of $\mathrm H$, the
Wigner--Eckart theorem immediately tells us that the last term of Eq.
\eqref{eq:two_derivative} vanishes. There is, however, a subtle exception to
this argument. As $\mathcal R$ is a complex representation, real
representations of $\mathrm H$ are doubled in its decomposition. The reason is
that when the set of vectors $\chi_k$ constitute the basis of a real
representation of $\mathrm H$, the vectors $i\chi_k$ form an independent basis
of an equivalent representation.

It may be that $H$ and $\Pi$ (or $\Pi\phi_0$) are such doubles. This happens,
for instance, for the two $5$-plets in Sec. III C. In such a case, however,
$\he\phi_0\partial_{\mu}\Pi\partial^{\mu}H$ is real and, again, does not
contribute to Eq. \eqref{eq:two_derivative}.

The single-derivative term in Eq. \eqref{eq:Lagrangian_sigma_model_expanded}
gives, after a short manipulation, the bilinear terms
\begin{equation}
-2\im\he HA^{\mu}\partial_{\mu}H-4\re\he
HA^{\mu}\partial_{\mu}\Pi\phi_0-\im\he\phi_0
A^{\mu}[\Pi,\partial_{\mu}\Pi]\phi_0. \label{eq:one_derivative}
\end{equation}
Throughout the calculation we made use of the fact that $A^{\mu}$ is a
$\mathrm{U(1)}$ generator, and therefore commutes with $\Pi$.

Putting together all the pieces of Eqs. \eqref{eq:two_derivative} and
\eqref{eq:one_derivative}, we arrive at our main result -- the bilinear
Lagrangian for a general linear sigma model,
\begin{multline}
\mathcal L_{\text{bilin}}=\partial_{\mu}\he
H\partial^{\mu}H-V_{\text{bilin}}(H)-2\im\he
HA^{\mu}\partial_{\mu}H\\
+\he\phi_0\partial_{\mu}\Pi\partial^{\mu}\Pi\phi_0-4\re\he
HA^{\mu}\partial_{\mu}\Pi\phi_0-\im\he\phi_0
A^{\mu}[\Pi,\partial_{\mu}\Pi]\phi_0. \label{eq:master_formula}
\end{multline}
This formula contains all information about the particle spectrum of the
theory, and the rest of the section is therefore devoted to its analysis.

\subsection{Discussion of the results}
There are altogether three terms with a single time derivative in Eq.
\eqref{eq:master_formula}. The term $\im\he HA^{\mu}\partial_{\mu}H$ causes
splitting of the masses of the massive modes. The term $\re\he
HA^{\mu}\partial_{\mu}\Pi\phi_0$ mixes massive and massless modes and,
according to Sec. II C produces linear GBs. Finally, the term $\im\he\phi_0
A^{\mu}[\Pi,\partial_{\mu}\Pi]\phi_0$ mixes the Goldstone fields and gives rise
to the quadratic Goldstones.

With the Wigner--Eckart theorem at hand it is easy to check that each of the
elementary fields appears in at most one of the three single-derivative terms.
This fact essentially reduces the analysis of the Lagrangian
\eqref{eq:master_formula} to the model two-field problem discussed in Sec. II
C.

To prove it note that the mixing term $\re\he HA^{\mu}\partial_{\mu}\Pi\phi_0$
can be nonzero only when $H$ and $\Pi$ are the two copies of the doubled real
representation of $\mathrm H$. Now the real multiplet $H$ gives real $\he
HA^{\mu}\partial_{\mu}H$, and therefore does not contribute to the mixing of
the massive modes. The real multiplet $\Pi$ analogously does not contribute to
$\im\he\phi_0 A^{\mu}[\Pi,\partial_{\mu}\Pi]\phi_0$ as a consequence of the
analysis that follows.

As the main concern of this paper is Goldstone boson counting, we shall now
concentrate on the last term of Eq. \eqref{eq:master_formula}, which produces
the quadratic GBs.

First, it is clear that our suspicion about the connection between the
quadratic GBs and nonzero charge densities was right. For by the very same
method as in Sec. III B we derive the Noether current corresponding to the
conserved charge $T$,
\begin{equation*}
j_T^{\mu}=-i(D^{\mu}\he\phi T\phi-\text{h.c.}),
\end{equation*}
and the ground-state density of $T$ is
\begin{equation}
j^0_T=2\he\phi_0A^0T\phi_0. \label{eq:charge_density}
\end{equation}
The last term of Eq. \eqref{eq:master_formula} is therefore indeed proportional to
the ground-state density of the commutator of two generators.

We may now in the general case proceed as in Sec. III, that is find the ground
state, calculate the Noether charge densities, and make a definite prediction for
the particle spectrum. We can, however, do even better, at least a bit.

We need not calculate the charge densities explicitly to say, which of the
generators may produce quadratic GBs. It is obvious from Eq.
\eqref{eq:charge_density} that only such a generator $T$ may acquire nonzero
density, which is a singlet of the unbroken subgroup $\mathrm{H}$. We therefore
just have to decompose the adjoint representation of $\mathrm{G}$ into
irreducible representations of $\mathrm{H}$ and look for spontaneously broken
singlets.

As in the examples above, we next choose such a basis that all the generators
with nonzero density mutually commute. This ensures that they can be completed
to form the Cartan subalgebra of the Lie algebra of $\mathrm G$. Following the
standard root decomposition of Lie algebras (see e.g. Ref.
\cite{Georgi:1982jb}), the rest of generators group into pairs whose commutator
lies in the Cartan subalgebra. They are the lowering and raising operators or
their hermitian linear combinations, and together with their commutator span an
$\mathrm{SU(2)}$ subalgebra of $\mathrm G$.

The point of this procedure is that only pairs of Goldstone fields are then
mixed by the single-derivative term $\im\he\phi_0
A^{\mu}[\Pi,\partial_{\mu}\Pi]\phi_0$ and the excitation spectrum may be fully
described with the help of the simple two-field bilinear Lagrangian
\eqref{eq:bilinear_Lagrangian}. Consequently, the quadratic GBs count as one
per each pair of generators whose commutator develops nonzero ground-state
density.

The feasibility of such a pairing also follows from group theory and the
Wigner--Eckart theorem. As the commutator of the two generators is to be an
$\mathrm{H}$-singlet, they must come from the same irreducible representation
of $\mathrm{H}$.

To briefly conclude this section, we once again emphasize the fact that almost
all we need to know about the excitation spectrum of the general linear sigma model
\eqref{eq:Lagrangian_sigma_model} may be extracted from the bilinear Lagrangian
\eqref{eq:master_formula} by simple group theory. We decompose the adjoint
representation of $\mathrm G$ with respect to the unbroken subgroup $\mathrm H$
to determine the multiplet structure of the Goldstones. The remaining $\mathrm
H$-multiplets in the decomposition of the representation $\mathcal R$ of the
scalar field $\phi$ are the massive modes.

The quadratic GBs are discovered with the knowledge of the ground-state
densities of the broken generators. Without further calculation, we can even
determine their dispersion relations. Making use of the continuity of the
dispersion relations across the phase transition and the known dispersion
relations in the unbroken phase, we may assert that the quadratic GB dispersion
relation is generically of the form $E=\vec p^2/2\mu Q$, where $Q$ is the
charge of the GB field under the $\mathrm{U(1)}$ subgroup equipped with the
chemical potential.

\section{Conclusions}
We have analyzed spontaneous breaking of internal symmetries in the framework
of the relativistic linear sigma model with finite chemical potential. Our
prime motivation was to establish a counting rule for Goldstone bosons in view
of the fact that explicit breaking of Lorentz invariance by medium effects may
cause the number of GBs to differ from the number of broken symmetry
generators.

Our results confirm the Nielsen--Chadha counting rule. We show
that the GBs have either linear or quadratic dispersion law at low momentum,
and that the number of the first plus twice the number of the second gives
exactly the number of broken generators.

In addition, we find a criterion which gives in a purely algebraic way the
number of quadratic GBs, the only necessary input being the structure of the
ground state. \emph{There is one quadratic GB for each pair of generators,
whose commutator has nonzero ground-state density.}

However, despite the generality of our results, many open questions still
remain. First, we stress the fact that we work all the time at the tree level.
It would be interesting to know the effect of radiative corrections on the
details of the spectrum. On the other hand, it seems that at least the
dispersion relations of the quadratic GBs are rather generic as they depend
only on the chemical potential in a very simple way. There might be a more
robust, nonperturbative method to determine them, which relies only on the
broken symmetry, and does not depend on the details of the dynamics of symmetry
breaking.

Second, we worked within the linear sigma model as it is easy to manipulate
perturbatively once the scalar field has been properly shifted to its new
ground state. It may happen that our results are valid generally for
relativistic theories with chemical potential. At least the argument presented
in Sec. II A that clarifies the connection between the charge densities and the
GB counting, suggests such a possibility.

As adding chemical potential breaks Lorentz invariance in a very particular
way, it might be possible to strengthen the Nielsen--Chadha counting rule at
the cost of limiting its validity to a smaller class of theories. Even such a
theorem would, however, find many applications on relativistic many-particle
systems. We hope that our future work will help to find the answer to these
questions.

\begin{acknowledgments}
The author is grateful to J. Ho\v{s}ek for critical reading of the manuscript
and numerous discussions. The present work was supported in part by the
Institutional Research Plan AV0Z10480505, and by the GACR grant No.
202/05/H003.
\end{acknowledgments}

\end{document}